\documentclass[twocolumn,showpacs,preprintnumbers,amsmath,amssymb]{revtex4}
\usepackage{epsfig}
\usepackage{graphicx}
\usepackage{dcolumn}
\usepackage{bm}
\usepackage{amsthm}
\usepackage{amsfonts}

\usepackage{color} 

\newcommand{\ket}[1]{\mbox{$ | #1 \rangle $}}
\newcommand{\bra}[1]{\mbox{$ \langle #1 | $}}

\begin{document}

\title{Upper bounds for the secure key rate of decoy state quantum key distribution}
\author{Marcos Curty$^1$, Tobias Moroder$^{2,3}$, Xiongfeng Ma$^2$, Hoi-Kwong Lo$^4$ and Norbert 
L\"{u}tkenhaus$^{2,3}$}
\affiliation{$^1$ ETSI Telecomunicaci\'on, Department of Signal Theory and Communications, University of Vigo, 
Campus Universitario, E-36310 Vigo (Pontevedra), Spain \\
$^2$ Institute for Quantum Computing, University of Waterloo, Waterloo, ON, N2L 3G1, Canada \\
$^3$ Quantum Information Theory Group, Institut f\"ur Theoretische
Physik I, and Max-Planck Research Group, Institute of Optics,
Information and Photonics, University of Erlangen-N\"urnberg,
91058 Erlangen, Germany \\
$^4$ Center for Quantum Information and Quantum Control (CQIQC), 
Department of Electrical \& Computer Engineering and Department of Physics,
University of Toronto, Toronto, ON, M5S 3G4, Canada}

\begin{abstract}
The use of decoy states in quantum key distribution (QKD) has provided a method for 
substantially increasing the secret key rate and distance that can be covered by QKD protocols 
with practical signals. The security analysis of these schemes, however, leaves open the possibility that 
the development of better proof techniques, or better classical post-processing methods, might 
further improve their performance in realistic scenarios. 
In this paper, we derive upper bounds on the secure key rate for decoy 
state QKD. These bounds are based basically only on the classical 
correlations established by the legitimate users during the quantum 
communication phase of the protocol. The only assumption about the 
possible post-processing methods is that double click events are 
randomly assigned to single click events. Further we consider only 
secure key rates based on the uncalibrated device scenario which assigns 
imperfections such as detection inefficiency to the eavesdropper. Our 
analysis
relies on two preconditions for secure two-way and one-way QKD:
The legitimate users need to prove that there exists no separable state (in the case of two-way QKD), or 
that there exists no quantum state having a symmetric extension (one-way QKD), that is compatible with the 
available measurements results. Both criteria have been previously applied to evaluate 
single-photon implementations of QKD. Here we use them to investigate a realistic source of weak coherent 
pulses. The resulting upper bounds can be formulated as a convex optimization 
problem known as a 
semidefinite program which can be efficiently solved. For the standard four-state QKD protocol, they 
are quite close to known lower bounds, 
thus showing that there are clear limits to the further improvement of classical post-processing techniques
in decoy state QKD. 

\end{abstract}

\maketitle

\section{Introduction}

Quantum key distribution (QKD) \cite{qkd,qkd1} allows two parties (Alice and Bob) to generate a secret key despite
the computational and technological power of an eavesdropper (Eve), who interferes with the signals. Together 
with the Vernam cipher \cite{vernam}, QKD can be used to
provide information-theoretic secure communications.

Practical QKD systems can differ in many important aspects from their original 
theoretical proposal, since these proposals typically demand technologies that are beyond our 
present experimental capability. Especially, the signals emitted by the source, instead of being single 
photons, are usually weak coherent pulses (WCP) with typical average photon numbers of 0.1 or higher. 
The quantum channel introduces errors and considerable attenuation (about $0.2$ dB/km) that affect the signals even when 
Eve is not present. Besides, for telecom wavelengths, standard InGaAs single-photon detectors
can have a detection 
efficiency below $15\%$ and are noisy due to dark counts. All these modifications jeopardize the security of the 
protocols, and lead to limitations 
of rate and distance that can be covered by these techniques \cite{pns}.

A main security threat of practical QKD schemes based on WCP arises from the fact that some signals contain 
more than one photon prepared in the same polarization state. Now Eve is no longer limited by the no-cloning 
theorem \cite{clon} since in these events the signal itself provides her with perfect copies of the signal photon.
She can perform, for instance, the so-called {\it photon number splitting} (PNS) attack on the multi-photon pulses
\cite{pns}. This attack gives Eve full information about the part of the key generated with the multi-photon 
signals, without causing any disturbance in the signal polarization. 
As a result, it turns out that the standard BB84 protocol \cite{BB84} with WCP can deliver a key generation rate of order
 $O(\eta^2)$, where $\eta$ denotes the transmission efficiency of the quantum channel \cite{opt,pracsecproofs1}.

To achieve higher secure key rates over longer distances, different QKD schemes, that are robust against the PNS 
attack, have been proposed in recent years \cite{decoy1,decoy2,decoy3,sarg,ben92}. 
One of these schemes is the so-called 
decoy state QKD \cite{decoy1,decoy2,decoy3} where 
Alice varies, independently and at random, the mean photon number of each signal state that she sends to 
Bob by employing different intensity settings.
Eve does not know a priori the mean photon number of each signal state sent by Alice. This means that 
her eavesdropping strategy can only depend on the photon number of these signals, but not on 
the particular intensity setting used to generate them. 
From the measurement results corresponding to different intensity settings, the legitimate users can 
estimate the classical joint probability distribution describing their outcomes for each photon number 
state. This provides 
them with a better estimation of the behaviour of the quantum channel, and it translates into an enhancement of the 
achievable secret key rate and distance. 
This technique has been successfully implemented in several recent experiments \cite{decoy_e}, 
and it can give a key generation rate of order $O(\eta)$ \cite{decoy1,decoy2,decoy3}. 

While the security analysis of decoy state QKD included in Refs.~\cite{decoy1,decoy2,decoy3}  is relevant from a 
practical point of view, it also 
leaves open the possibility that the development of better proof techniques, or better classical post-processing 
protocols, might further improve the performance of these schemes in realistic scenarios.
For instance, it is known that two-way classical post-processing protocols can tolerate a 
higher error rate than one-way communication techniques \cite{two_way,two_way2}, 
or that by modifying the public announcements of the standard BB84 protocol it is possible to generate a secret key 
even from multi-photon signals \cite{sarg}.
Also, the use of local randomization \cite{local_rand} and degenerate codes \cite{deg} 
can as well improve the error rate thresholds of the protocols. 

In this paper we 
consider the uncalibrated device scenario \cite{qkd1} and we assume the typical initial post-processing 
step where double click events are not discarded by Bob, but they are randomly assigned to 
single click events \cite{ndc}. In this scenario, we derive
simple upper bounds on the secret key rate and distance that 
can be covered by decoy state 
QKD based 
exclusively
on the classical correlations established by the legitimate users
during the quantum communication phase of the protocol. 
Our analysis relies on two preconditions for secure two-way and one-way QKD. In particular, 
Alice and Bob need to prove that there exists no separable state (in the case of two-way QKD) 
\cite{curty04a,curty04b}, or that there exists no quantum state having a symmetric extension (one-way QKD)
\cite{tobione}, that is compatible with the available measurements results. 
Both criteria have been already applied 
to evaluate single-photon implementations of 
QKD \cite{curty04a,curty04b,tobione,tobi_two,tobithree}. Here we employ them for the first time to investigate 
practical realizations of QKD based on the distribution of WCP. 

We show that
both preconditions for secure two-way and one-way QKD can be formulated as a convex optimization 
problem known as a 
semidefinite program (SDP) \cite{sdp}. Such instances of convex optimization problems 
appear frequently in quantum information theory and can be solved with arbitrary accuracy
in polynomial time, for example, by the interior-point methods \cite{sdp}. 
As a result, we obtain ultimate upper bounds on the performance of decoy state QKD 
when this typical initial post-processing of the double clicks is performed.
These upper bounds hold for any possible 
classical communication
technique that  the legitimate users 
can employ 
in this scenario afterwards
like, for example, the SARG04 protocol \cite{sarg}, adding noise protocols \cite{local_rand}, 
degenerate codes protocols \cite{deg} and 
two-way classical post-processing protocols \cite{two_way,two_way2}.
The analysis 
presented in this manuscript can as well be straightforwardly adapted to evaluate
other implementations of the BB84 protocol with practical signals as, for instance, 
those experimental demonstrations based on  WCP without 
decoy states or on entangled signals coming from a parametric down conversion source. 

The paper is organized as follows. In Sec.~\ref{sec_a} we describe in detail a 
WCP implementation of the BB84 protocol based on decoy states. Next, in Sec.~\ref{sec_b} we apply 
two criteria for secure two-way and one-way QKD to this scenario.
Here we derive upper bounds on the secret key rate and distance that 
can be achieved with decoy state QKD as a function of the observed quantum bit error rate (QBER) 
and the losses in the quantum channel. Moreover, we show how to cast both upper bounds 
as SDPs. These results are then illustrated in Sec.~\ref{sec_c} for the case of  
a typical behaviour of the quantum channel, {\it i.e.}, in the 
absence of eavesdropping. Finally, Sec.~\ref{CONC} concludes the paper with a summary.

\section{Decoy state QKD}
\label{sec_a}

In decoy state QKD with WCP Alice prepares phase-randomized coherent states with 
Poissonian photon number distribution. The mean photon number 
(intensity) of this distribution is chosen at random for each signal from a set of 
possible values $\mu_l$.
In the case of the BB84 protocol, 
and assuming Alice chooses a decoy intensity setting $l$,
such states can be described as
\begin{equation}\label{sig1}
\rho_B^k(\mu_l)=e^{-\mu_l}\sum_{n=0}^{\infty} \frac{\mu_l^n}{n!}\ket{n_k}_B\bra{n_k},
\end{equation}
where the signals $\ket{n_k}_B$ denote Fock states with $n$ photons in one of the four possible polarization 
states of the BB84 scheme, which are labeled with the index $k\in\{0, \ldots, 3\}$. On the 
receiving side, we consider that Bob employs an active basis choice measurement 
setup. This device splits the incoming light by means of a polarizing beam-splitter and 
then sends it to threshold detectors that cannot resolve the number of photons by which they are
triggered. The polarizing beam-splitter can be oriented along any of the 
two possible polarization 
basis used in the BB84 protocol. This detection setup is characterized by one {\it positive operator value 
measure} (POVM) that we shall denote as
$\{B_j\}$. 

In an entanglement-based view, the signal preparation process described above can 
be modeled 
as follows: Alice produces first bipartite states of the form
\begin{equation}\label{mcl}
\ket{\Psi_{\rm source}}_{AB}=\sum_{k=0}^{3}\sum_{l=0}^{\infty} \sqrt{q_k p_l}
\ket{k}_{A_1}\ket{l}_{A_2}\ket{\phi_{kl}}_{A_3B}, 
\end{equation}
where system $A$ is the composition of systems $A_1$, $A_2$, and $A_3$, and
the orthogonal states $\ket{k}_{A_1}$ and $\ket{l}_{A_2}$
record, respectively, the polarization state and decoy intensity setting selected by Alice. 
The parameters
$q_k$ and $p_l$ represent the a priori probabilities of these signals. For instance, in 
the standard BB84 scheme the four possible polarization states are chosen with equal a
priori probabilities and $q_k=1/4$ for all $k$.
The signal 
$\ket{\phi_{kl}}_{A_3B}$ that appears 
in Eq.~(\ref{mcl})
denotes a 
purification of the state $\rho_B^k(\mu_l)$ and can be written as
\begin{equation}\label{pur_fnl}
\ket{\phi_{kl}}_{A_3B}=e^{-\mu_l/2}\sum_{n=0}^{\infty} 
\frac{\sqrt{\mu_l}^n}{\sqrt{n!}}\ket{n}_{A_3}\ket{n_k}_B,
\end{equation}
where 
system $A_3$ acts as a shield, in the sense of Ref.~\cite{shield} and
records the photon number information 
of the signals prepared by the source. This system
is typically inaccessible to all the parties.
One could also select 
as $\ket{\phi_{kl}}_{A_3B}$ any other purification of the state $\rho_B^k(\mu_l)$. 
However, as we will show in Sec.~\ref{sec_b}, the one given by Eq.~(\ref{pur_fnl}) is particularly 
suited for the calculations that we present in that section.

Afterwards, Alice measures systems $A_1$ and $A_2$ in the orthogonal basis 
$\ket{k}_{A_1}$ and $\ket{l}_{A_2}$, 
corresponding to the measurement operators 
$A_{kl}=\ket{k}_{A_1}\bra{k}\otimes\ket{l}_{A_2}\bra{l}$. 
This action generates the signal states 
$\rho_B^k(\mu_l)$ with a priori probabilities $q_k p_l$. The reduced density matrix
$\rho_{A}={\rm Tr}_{B}(\rho_{AB})$,
with $\rho_{AB}=\ket{\Psi_{\rm source}}_{AB}\bra{\Psi_{\rm source}}$,
is fixed by the actual preparation scheme and cannot be modified by Eve. 
In order to include this information in the 
measurement process, one can add to the observables $\{A_{kl}\otimes{}B_j\}$, 
measured by Alice and Bob, other observables  
$\{C_{i}\otimes\openone_B\}$ such that $\{C_{i}\}$ form a 
complete tomographic set of Alice's 
Hilbert space $\mathcal{H}_{A}$
\cite{curty04b}. In order to simplify our notation, from now on we shall consider that the observed data
$p_{klj}={\rm Tr}(A_{kl}\otimes{}B_j\ \rho_{AB})$ and the 
POVM $\{A_{kl}\otimes{}B_j\}$ contain also the observables 
$\{C_{i}\otimes\openone_B\}$. That is, every time we refer to $\{A_{kl}\otimes{}B_j\}$
we assume that these operators include as well the observables $\{C_{i}\otimes\openone_B\}$. 

\section{Upper bounds on decoy state QKD}
\label{sec_b}

Our starting point is the observed joint probability distribution
$p_{klj}$ obtained by Alice and Bob after their measurements
$\{A_{kl}\otimes B_j\}$. This probability distribution
defines an equivalence class $\mathcal{S}$ of quantum states that are
compatible with it,
\begin{equation}
  \label{eq_class}
  \mathcal{S}=\big\{ \sigma_{AB}\ |\ \text{Tr}(A_{kl} \otimes B_j\
  \sigma_{AB})=p_{klj} \ \forall k,l,j \big\}.
\end{equation}

\subsection{Two-way classical post-processing}\label{sec_twoway}

Let us begin by considering two-way classical post-processing of the data  $p_{klj}$. 
It was shown in Ref.~\cite{curty04b} that a necessary precondition to distill a secret key in this scenario
is that the equivalence class $\mathcal{S}$ does not contain any separable state. 
That is, we need to find quantum-mechanical correlations 
in $p_{klj}$, otherwise the secret key rate, that we shall denote as $K$, vanishes
\cite{nwnt}. As it is, this precondition 
answers only partially the important question of how much secret key Alice and Bob can
obtain from their correlated data. It just tells if the secret key rate is zero or it may be 
positive. However, this criterion can be used as a benchmark to evaluate any 
upper bound on $K$. If $\mathcal{S}$ contains a separable state then the upper bound 
must vanish. One upper bound which satisfies this condition is that given 
by the regularized relative entropy of entanglement \cite{up1}.
Unfortunately, to calculate this quantity for a given quantum 
state is, in general, a quite difficult task, and analytical expressions are only available 
for some particular cases \cite{regent2}. Besides, this upper bound depends 
exclusively on the quantum states shared by Alice and Bob and, therefore, it does not 
include 
the effect of imperfect devices like, for instance, the low detection efficiency 
or the noise in the form of dark counts introduced by current detectors \cite{tobi_two}.
Another possible approach is that 
based on the best separable approximation (BSA) of a quantum state
$\sigma_{AB}$ \cite{BSA}. This is the decomposition of $\sigma_{AB}$ into a 
separable state $\sigma_{sep}$ and an entangled state $\rho_{ent}$, while maximizing the 
weight of the separable part. That is, any quantum state
$\sigma_{AB}$ can always be written as
\begin{equation}\label{eq_BSA}
\sigma_{AB}=\lambda(\sigma_{AB})\sigma_{sep}+[1-\lambda(\sigma_{AB})]\rho_{ent},
\end{equation}
where the real parameter $\lambda(\sigma_{AB})\geq{}0$ is maximal.

Given an equivalence class $\mathcal{S}$ of quantum states, one can define the maximum 
weight of separability within the class, $\lambda_{BSA}^\mathcal{S}$, as
\begin{equation}
\lambda_{BSA}^\mathcal{S}=\text{max}\{\lambda(\sigma_{AB})\ |\  \sigma_{AB}\in\mathcal{S}\}.
\end{equation}
Note that the correlations $p_{klj}$ can originate from a separable state if and only if $\lambda_{BSA}^\mathcal{S}=1$.
Let $\mathcal{S}_{BSA}^{ent}$ denote the equivalence class of quantum
states given by 
\begin{equation}
\mathcal{S}_{BSA}^{ent}=\{\rho_{ent}\ |\  \sigma_{AB}\in\mathcal{S}\ \text{and}\
\lambda(\sigma_{AB})=\lambda_{BSA}^\mathcal{S} \},
\end{equation}
where $\rho_{ent}$ represents again the entangled part in the BSA of the state $\sigma_{AB}$. 
Then, it was proven in Ref.~\cite{tobi_two} that the secret key rate $K$ always satisfies
\begin{equation}\label{second_bound}
K\leq{}(1-\lambda_{BSA}^\mathcal{S})I^{ent}(A;B), 
\end{equation}  
where $I^{ent}(A;B)$ represents the Shannon mutual information calculated 
on the joint probability distribution $q_{klj}=\text{Tr}(A_{kl} \otimes B_j\ \rho_{ent})$.
As it is, this upper bound can be applied to any QKD scheme \cite{tobi_two}, although the
calculation of the parameters $\lambda_{BSA}^\mathcal{S}$ and $\rho_{ent}$ might be a challenge.
Next, we consider the particular case of 
decoy state QKD.

\subsubsection*{Upper bound on two-way decoy state QKD}

The signal states $\rho_B^k(\mu_l)$ that Alice sends to Bob are mixtures of Fock states 
with different Poissonian photon number distributions of mean $\mu_l$. 
This means, in particular, that Eve can always perform a {\it quantum non-demolition} (QND) measurement of the 
total number of photons contained in each of these signals
without introducing 
any errors. The justification 
for this is that the total photon number information via the QND measurement ``comes free", since the execution 
of this measurement does not change the signals $\rho_B^k(\mu_l)$.
That is, the realization of this measurement cannot make Eve's eavesdropping capabilities weaker \cite{usda}. 
If Eve 
performs such a QND measurement, then the signals $\rho_{AB}=\ket{\Psi_{\rm source}}_{AB}\bra{\Psi_{\rm source}}$ 
are transformed as
\begin{equation}\label{trans}
\rho_{AB} \mapsto \gamma_{AB}
=\sum_{n=0}^{\infty} r_n \ket{\varphi_n}_{A_1B}\bra{\varphi_n}\otimes
\ket{\mu_n}_{A_2}\bra{\mu_n}\otimes
\ket{n}_{A_3}\bra{n},
\end{equation} 
where 
the probabilities $r_n$ are given by
\begin{equation}\label{rn}
r_n=\sum_{l=0}^{\infty} p_l \frac{e^{-\mu_l}\mu_l^n}{n!},
\end{equation}
the signals  
$\ket{\varphi_n}_{A_1B}$ have the form
\begin{equation}\label{eq_nsq}
\ket{\varphi_n}_{A_1B}=\sum_{k=0}^{3} \sqrt{q_k} \ket{k}_{A_1}\ket{n_k}_B,
\end{equation}
and the normalized states $\ket{\mu_n}_{A_2}$ only depend on 
the signals $\ket{l}_{A_2}$ and the photon number $n$.

From the tensor product structure of $\gamma_{AB}$ we learn that 
the signals $\gamma_{AB}$ can only contain quantum correlations between systems $A_1$ and 
$B$.
Therefore, without loss of generality, we can always restrict ourselves to only search for quantum correlations 
between these two systems. 
Additionally, in decoy state QKD Alice and Bob have always access to 
the conditional joint probability distribution describing their outcomes given that 
Alice emitted an $n$-photon state. This means that the search for quantum correlations in 
$\mathcal{S}$
can be done independently for each $n$-photon signal. That is, the legitimate users can 
define an equivalence class of signal states for each possible Fock state sent by Alice. 

A further simplification arises when one
considers the 
typical initial post-processing 
step where double click events are not discarded by Bob, but they are randomly assigned to 
single click events \cite{ndc}. In the case of the BB84 protocol, this action
allows Alice and Bob to always explain their observed data as coming from a 
single-photon signal where Bob performs a single-photon measurement $\{T_j\}$ \cite{squash}. 
This measurement is characterized by a set of POVM operators which are projectors
onto the eigenvectors
of the two Pauli operators $\sigma_x$ and $\sigma_z$, together with a projection onto the 
vacuum state $\ket{vac}$ which models the losses in the quantum channel, 
\begin{eqnarray}\label{op_Tj} 
T_0&=&\frac{1}{2}\ket{0}_B\bra{0},  \quad\quad T_1=\frac{1}{2}\ket{1}_B\bra{1}, \nonumber\\ 
T_\pm&=&\frac{1}{2}\ket{\pm}_B\bra{\pm}, \quad\quad T_{vac}=\ket{vac}_B\bra{vac}, 
\end{eqnarray}
with $\ket{\pm}=(\ket{0}\pm\ket{1})/\sqrt{2}$ and where $\sum_j T_j=\openone_B$ \cite{squash}.
In particular, let $p_{kj}^n$ denote the conditional joint probability distribution obtained 
by Alice and Bob after their measurements $\{A_k\otimes{}T_j\}$,
with $A_k=\ket{k}_{A_1}\bra{k}$, 
given that 
Alice emitted an $n$-photon state. That is, $p_{kj}^n$ includes the random 
assignment of double clicks to single click events.  
As before, we consider that the observables 
$\{A_k\otimes{}T_j\}$ contain as well other observables $\{C_i\otimes\openone_B\}$ that
form a tomographic complete set of Alice's Hilbert space $\mathcal{H}_{A_1}$. 
We define the 
equivalence class $\mathcal{S}^n$ of quantum states 
that are compatible with $p_{kj}^n$ as
\begin{equation}
  \label{eq_class_n_squash}
  \mathcal{S}^n=\big\{ \sigma_{A_1B}^{n}\ |\ \text{Tr}(A_{k} \otimes T_j\
  \sigma_{A_1B}^{n})=p_{kj}^n, \ \forall k,j \}.
\end{equation}
Then, the secret key rate $K$ can be upper bounded as
\begin{equation}\label{final_two}
K\leq{}\sum_{n\geq{}1}  r_n (1-\lambda_{BSA}^{\mathcal{S}^n})I_{n}^{ent}(A;B),
\end{equation}  
where $\lambda_{BSA}^{\mathcal{S}^n}$ denotes the maximum weight of separability within the 
equivalence class $\mathcal{S}^n$, and $I_{n}^{ent}(A;B)$
represents the
Shannon mutual information calculated on
$q_{kj}^{n}=\text{Tr}(A_{k} \otimes T_j\ \rho_{ent}^{n})$, with
$\rho_{ent}^{n}$ being the entangled part in the BSA of a state $\sigma_{A_1B}^{n}\in\mathcal{S}^n$ and whose 
weight of separability is maximum. 

The main difficulty when evaluating the upper bound given by Eq.~(\ref{final_two}) still relies 
on obtaining the parameters $\lambda_{BSA}^{\mathcal{S}^n}$ and $\rho_{ent}^{n}$. 
Next, we show how to solve this problem by means of a semidefinite program (SDP) \cite{sdp}. 
For that, we need to prove first the 
following observation. 

{\it Observation}: Within the equivalence classes $\mathcal{S}^n$ of quantum signals 
given by Eq.~(\ref{eq_class_n_squash})
Alice and Bob can only 
detect the presence of negative partial transposed (NPT) entangled states \cite{NPT}.

{\it Proof}. 
The signals $\sigma_{A_1B}^{n}\in \mathcal{S}^n$ can always be decomposed as
\begin{equation}\label{form_ss}
\sigma_{A_1B}^{n}=p\tilde{\rho}_{A_1B}^n+(1-p)\tilde{\rho}_{A_1}^n\otimes\ket{vac}_B\bra{vac},
\end{equation}
for some probability $p\in[0,1]$, and
where $\tilde{\rho}_{A_1B}^n\in\mathcal{H}_{A_1}\otimes\mathcal{H}_{2}$, and 
$\tilde{\rho}_{A_1}^n\in\mathcal{H}_{A_1}$.
That is, the state 
$\sigma_{A_1B}^{n}$ can only be entangled if $\tilde{\rho}_{A_1B}^n$ is also entangled. 
In order to detect entanglement in 
the latter one,
Bob projects it 
onto the eigenvectors of the two Pauli operators 
$\sigma_x$ and $\sigma_z$. This means, in particular, that the class of {\it accessible} entanglement 
witness operators $W$ that can be constructed from the available measurements results satisfy 
$W=W^{\Gamma}$. Here $\Gamma$ denotes transposition 
with respect to Bob's system. We have, therefore, that 
$\text{Tr}(W \tilde{\rho}_{A_1B}^{n})=\text{Tr}(W \Omega)$, with
$\Omega=\frac{1}{2}[\tilde{\rho}_{A_1B}^{n}+\tilde{\rho}_{A_1B}^{n\ \Gamma}]$.
For the given dimensionalities, it 
was proven in Ref.~\cite{kckl} that whenever $\Omega$ is non-negative
it represents a separable state, {\it i.e.}, $\text{Tr}(W \Omega)\geq{}0$.
This means that Alice and Bob can only detect entangled states 
$\tilde{\rho}_{A_1B}^n$ that 
satisfy $\Omega\ngeq{}0$. Since $\tilde{\rho}_{A_1B}^{n}\geq{}0$, the previous 
condition is only possible when $\tilde{\rho}_{A_1B}^{n\ \Gamma}\ngeq{}0$.
$\blacksquare$

Let us now write the search of $\lambda_{BSA}^{\mathcal{S}^n}$ and 
$\rho_{ent}^{n}$ as a SDP. This is a convex optimisation problem
of the following form \cite{sdp}:
\begin{eqnarray}\label{primalSDP_marcos}
\text{minimize} && c^T {\bf{x}} \\
\nonumber \text{subject to} && F({\bf{x}})=F_0 + \sum_i x_i F_i
\geq 0,
\end{eqnarray}
where the vector ${\bf x}$ represents the
objective variable, the vector $c$ is fixed by the particular
optimisation problem, and the matrices $F_0$ and $F_i$ are
Hermitian matrices. The goal is to minimize the linear function
$c^T{\bf{x}}$ subjected to the linear matrix inequality (LMI)
constraint $F({\bf{x}}) \geq 0$. The SDP that we need to solve 
has the form \cite{aclar_sdp}:
\begin{eqnarray}\label{primalSDP_two_way}
\text{minimize} && 1-\text{Tr}[\sigma_{sep}^n({\bf x})] \\
\nonumber \text{subject to} && \sigma_{A_1B}^n({\bf x})\geq{}0, \\
\nonumber && \text{Tr}[\sigma_{A_1B}^n({\bf x})]=1, \\
\nonumber && \text{Tr}[A_{k} \otimes T_j\ \sigma_{A_1B}^n({\bf x})]=p_{kj}^n, \ \forall k,j, \\
\nonumber && \sigma_{sep}^n({\bf x})\geq{}0, \\
\nonumber && \sigma_{sep}^{n\ \Gamma}({\bf x})\geq{}0, \\
\nonumber && \sigma_{A_1B}^n({\bf x})-\sigma_{sep}^n({\bf x})\geq{}0, 
\end{eqnarray}
where the objective variable ${\bf x}$ is used to parametrise the
density operators $\sigma_{sep}^n$ and $\sigma_{A_1B}^n$. 
For that, we employ the method 
introduced in Refs.~\cite{tobi_two,tobithree}.
The state $\sigma_{sep}^n$ which appears in Eq.~(\ref{primalSDP_two_way})
is not normalized, {\it i.e.}, it also includes the parameter 
$\lambda(\sigma_{A_1B}^n)$. The first three constraints in Eq.~(\ref{primalSDP_two_way})
guarantee that $\sigma_{A_1B}^n$ is a valid normalized density operator that 
belongs to the equivalence class $\mathcal{S}^n$, the following two 
constraints impose $\sigma_{sep}^n$ to be a separable state, while the last 
one implies that the entangled part of $\sigma_{A_1B}^n$ is a valid but not normalized density 
operator. Its normalization factor is given by $1-\lambda(\sigma_{A_1B}^n)$. 
If ${\bf x}_{sol}$ denotes a 
solution to the SDP given by Eq.~(\ref{primalSDP_two_way}) then  
\begin{equation}
\lambda_{BSA}^{\mathcal{S}^n}=\text{Tr}[\sigma_{sep}^n({\bf x}_{sol})],
\end{equation}
and the state $\rho_{ent}^{n}$ is given by
\begin{equation}
\rho_{ent}^{n}=\frac{\sigma_{A_1B}^n({\bf x}_{sol})-\sigma_{sep}^n({\bf x}_{sol})}{1-\lambda_{BSA}^{\mathcal{S}^n}}.
\end{equation}

\subsection{One-way classical post-processing}\label{sec_one_way}

The classical post-processing of the observed data can 
be restricted to 
one-way
communication \cite{initial_2w}. 
Depending on the allowed direction of 
communication, two different cases can be considered: {\it Direct reconciliation} (DR) 
refers to communication from Alice to Bob, {\it reverse reconciliation} (RR) permits only communication 
from Bob to Alice \cite{rrdr}. In this section, we will only consider the case of DR. 
Expressions for the opposite scenario, {\it i.e.}, RR, can be obtained in a similar way. 
In Ref.~\cite{tobione} it was shown that a necessary precondition for secure 
QKD by means of DR (RR) is that the equivalence class $\mathcal{S}$ 
given by Eq.~(\ref{eq_class})
does not contain any state having a symmetric extension to two copies of system $B$ (system $A$). 

\begin{figure}
\begin{center}
\includegraphics[angle=0,scale=.7]{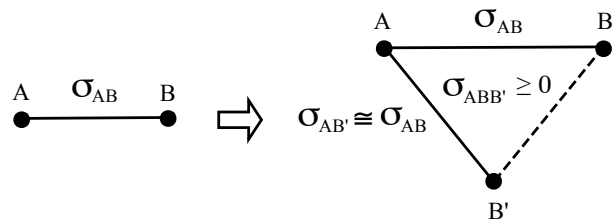}
\end{center}
\caption{Graphical illustration of a 
quantum state 
$\sigma_{AB}$ which has a 
symmetric extension to two copies of system $B$. \label{fig_sym}}
\end{figure}

A state $\sigma_{AB}$ is said to have a symmetric extension to two copies of system $B$ if and only if 
there exists a tripartite state $\sigma_{ABB^\prime} \geq 0$, with
$\text{Tr}(\sigma_{ABB^\prime})=1$, and where $\mathcal{H}_B\cong\mathcal{H}_{B^\prime}$, 
which fulfills the following two properties \cite{do1}:
\begin{eqnarray}
\textrm{Tr}_{B^\prime}(\sigma_{ABB^\prime}) & = & \sigma_{AB},\label{eq1}\\
P \sigma_{ABB^\prime} P & = & \sigma_{ABB^\prime},\label{eq2}
\end{eqnarray}
where the swap operator $P$ satisfies $P \ket{ijk}_{ABB^\prime} =
\ket{ikj}_{ABB^\prime}$. A graphical illustration of a state $\sigma_{AB}$ which has a 
symmetric extension to two copies of system $B$ is given in Fig.~\ref{fig_sym}. 
This definition can be easily extended
to cover also the case of symmetric extensions of $\sigma_{AB}$ to
two copies of system $A$, and also of extensions of $\sigma_{AB}$
to more than two copies of system $A$ or of system $B$.

The best extendible approximation (BEA) of a given state $\sigma_{AB}$ is the decomposition of 
$\sigma_{AB}$ into a state with a symmetric extension, that we denote as 
$\sigma_{ext}$, and a state without symmetric extension $\rho_{ne}$, 
while maximizing the weight of the extendible part, {\it i.e.},
\begin{equation}\label{blia}
\sigma_{AB}=\lambda(\sigma_{AB})\sigma_{ext}+[1-\lambda(\sigma_{AB})]\rho_{ne},
\end{equation}
where the real parameter $\lambda(\sigma_{AB})\geq{}0$ is maximal \cite{tobione,Note_sym}. 
Note that this parameter is well defined since the set of extendible states is 
compact.

Equation~(\ref{blia})
follows the same spirit like the BSA given by Eq.~(\ref{eq_BSA}). Now, one can define 
analogous parameters and equivalence classes as in 
Sec.~\ref{sec_twoway}. In particular, the maximum weight of extendibility within 
an equivalence class $\mathcal{S}$ is defined as
$\lambda_{BEA}^\mathcal{S}=\text{max}\{\lambda(\sigma_{AB})\ |\  \sigma_{AB}\in\mathcal{S}\}$.
That is, the correlations $p_{klj}=\text{Tr}(A_{kl} \otimes B_j\ \sigma_{AB})$ can originate from an
extendible state if and only if $\lambda_{BEA}^\mathcal{S}=1$. Finally, one defines
$\mathcal{S}_{BEA}^{ne}$ as the equivalence class of quantum
states given by 
$\mathcal{S}_{BEA}^{ne}=\{\rho_{ne}\ |\  \sigma_{AB}\in\mathcal{S}\ \text{and}\
\lambda(\sigma_{AB})=\lambda_{BEA}^\mathcal{S} \}$,
where $\rho_{ne}$ denotes the nonextendible part in the BEA of the state $\sigma_{AB}$. 
Then, it was proven in Ref.~\cite{tobione}
that the one-way secret key rate $K_{\rightarrow}$ satisfies 
\begin{equation}\label{first_bound_one}
K_{\rightarrow}\leq{}(1-\lambda_{BEA}^\mathcal{S})I^{ne}(A;B),
\end{equation}
where $I^{ne}(A;B)$ represents the Shannon mutual information now calculated 
on the joint probability distribution $q_{klj}=\text{Tr}(A_{kl} \otimes B_j\ \rho_{ne})$ with 
$\rho_{ne}\in\mathcal{S}_{BEA}^{ne}$. 

\subsubsection*{Upper bound on one-way decoy state QKD}

The analysis contained in Sec.~\ref{sec_twoway} to derive Eq.~(\ref{final_two}) 
from Eq.~(\ref{second_bound})
also applies to this scenario and we omit it here for simplicity. We obtain
\begin{equation}\label{final_one}
K_{\rightarrow}\leq{}\sum_{n\geq{}1}  r_n (1-\lambda_{BEA}^{\mathcal{S}^n})I_{n}^{ne}(A;B).
\end{equation}  
where $\lambda_{BEA}^{\mathcal{S}^n}$ denotes the maximum weight of 
extendibility within the equivalence class $\mathcal{S}^n$ given by 
Eq.~(\ref{eq_class_n_squash}), 
and $I_{n}^{ne}(A;B)$ represents the Shannon mutual information calculated
on $q_{kj}^{n}=\text{Tr}(A_k\otimes{}T_j\ \rho_{ne}^{n})$, with $\rho_{ne}^{n}$ being 
the nonextendible part in the BEA of a state 
$\sigma_{A_1B}^{n}\in\mathcal{S}^n$ and whose weight of extendibility is 
maximum. 

The parameter $\lambda_{BEA}^{\mathcal{S}^n}$ and the nonextendible state 
$\rho_{ne}^{n}$ can directly be obtained by solving the following SDP: 
\begin{eqnarray}\label{primalSDP_one_way}
\text{minimize} && 1-\text{Tr}[\sigma_{ext}^n({\bf x})] \\
\nonumber \text{subject to} && \sigma_{A_1B}^n({\bf x})\geq{}0, \\
\nonumber && \text{Tr}[\sigma_{A_1B}^n({\bf x})]=1, \\
\nonumber && \text{Tr}[A_{k} \otimes T_j\ \sigma_{A_1B}^n({\bf x})]=p_{kj}^n, \ \forall k,j, \\
\nonumber && \rho_{A_1BB'}^n({\bf x})\geq{}0, \\
\nonumber && P\rho_{A_1BB'}^n({\bf x})P=\rho_{A_1BB'}^n({\bf x}), \\
\nonumber && {\rm Tr}_{B'}[\rho_{A_1BB'}^n({\bf x})]=\sigma_{ext}^n({\bf x}), \\
\nonumber && \sigma_{A_1B}^n({\bf x})-\sigma_{ext}^n({\bf x})\geq{}0, 
\end{eqnarray}
where 
the state $\sigma_{ext}^n$ 
is not normalized, {\it i.e.}, it also includes the parameter 
$\lambda(\sigma_{A_1B}^n)$. The first three constraints 
coincide with those of Eq.~(\ref{primalSDP_two_way}). 
They just guarantee that $\sigma_{A_1B}^n\in\mathcal{S}^n$. The following three
constraints impose $\sigma_{ext}^n$ to have a symmetric extension to 
two copies of system $B$, while the last one
implies that the nonextendible 
part of $\sigma_{A_1B}^n$ is a valid but not normalized density 
operator. Its normalization factor is $1-\lambda(\sigma_{A_1B}^n)$. 
This SDP does not include the constraint $\sigma_{ext}^n \geq 0$ because non-negativity 
of the extension $\rho_{A_1BB'}^n$, together with the condition
$\text{Tr}_{B'}(\rho_{A_1BB'}^n)=\sigma_{ext}^n$, already implies non-negativity of $\sigma_{ext}^n$. 
If ${\bf x}_{sol}$ represents a 
solution to the SDP given by Eq.~(\ref{primalSDP_one_way}) then 
we have that
\begin{equation}
\lambda_{BEA}^{\mathcal{S}^n}=\text{Tr}[\sigma_{ext}^n({\bf x}_{sol})],
\end{equation} 
and the state $\rho_{ne}^{n}$ is given by
\begin{equation}
\rho_{ne}^{n}=\frac{\sigma_{A_1B}^n({\bf x}_{sol})-\sigma_{ext}^n({\bf x}_{sol})}{1-\lambda_{BEA}^{\mathcal{S}^n}}.
\end{equation}

\section{Evaluation}
\label{sec_c}

In this section we evaluate the upper bounds on the secret key rate both for two-way and one-way 
decoy state QKD given by Eq.~(\ref{final_two}) and Eq.~(\ref{final_one}).
Moreover, we compare our results with known lower 
bounds for the same scenarios. The numerical simulations
are performed with the freely available SDP solver SDPT3-3.02 \cite{sdpt3}, together with the 
parser YALMIP \cite{yalmip}. 

\subsection{Channel model}

To generate the observed data, we consider the channel model used in Ref.~\cite{decoy2,phdxiongfeng}. 
This model reproduces a normal behaviour 
of the quantum channel, {\it i.e.}, in the absence of eavesdropping. 
Note, however, that
our analysis can as well be
straightforwardly applied to other quantum channels, as it only depends on the probability distribution 
$p_{kj}^n$ that characterizes the results of Alice's and Bob's measurements.
This probability distribution 
is given in Tab.~\ref{tab1}, where the conditional
yields $Y_n$ have the form
\begin{table}
\begin{tabular}{|c|c|c|c|c|c|}
\hline\hline $p_{kj}^n$
\rule{0mm}{3.9mm}& $T_{j=0}$ & $T_{j=1}$ & $T_{j=+}$ &
$T_{j=-}$ &
    $T_{j=vac}$ \\
\hline
$k=0$\rule{0mm}{3.5mm}& $\frac{Y_n(1-e_n)}{8}$ & $\frac{Y_ne_n}{8}$ & $\frac{Y_n}{16}$ & $\frac{Y_n}{16}$ & $\frac{1-Y_n}{4}$ \\
$k=1$& $\frac{Y_ne_n}{8}$ & $\frac{Y_n(1-e_n)}{8}$ & $\frac{Y_n}{16}$ & $\frac{Y_n}{16}$ & $\frac{1-Y_n}{4}$ \\
$k=2$& $\frac{Y_n}{16}$ & $\frac{Y_n}{16}$ & $\frac{Y_n(1-e_n)}{8}$ & $\frac{Y_ne_n}{8}$ & $\frac{1-Y_n}{4}$ \\
$k=3$& $\frac{Y_n}{16}$ & $\frac{Y_n}{16}$ & $\frac{Y_ne_n}{8}$ & $\frac{Y_n(1-e_n)}{8}$ & $\frac{1-Y_n}{4}$  \\
\hline\hline
\end{tabular}
\caption{Conditional joint probability distribution
$p_{kj}^n=\text{Tr}(A_k \otimes T_j\ \sigma_{A_1B}^{n})$, where the index $k\in\{0,\ldots,3\}$ labels, 
respectively, the 
four possible polarization states of the BB84 protocol ($0,1,+,-$), and the operators $T_j$ are given by Eq.~(\ref{op_Tj}).  
It satisfies $\sum_{k,j} p_{kj}^n=1$.}\label{tab1}
\end{table}
\begin{equation}
Y_n=Y_0+[1-(1-\eta)^n],
\end{equation}
with $Y_0$ being the background detection event rate of the system, and where $\eta$ represents the overall 
transmittance, including the transmission efficiency of the quantum channel and the detection efficiency.
The parameter $e_n$ denotes the quantum bit error rate of an $n$-photon signal. It is 
given by
\begin{equation}
e_n=\frac{e_{det}[1-(1-\eta)^n]+\frac{1}{2}Y_0}{Y_n},
\end{equation}
where $e_{det}$ represents the probability that a photon hits the wrong detector due to the 
misalignment in the quantum channel and in the detection apparatus. 

The parameter $\eta$ can be related with a transmission distance $l$ measured in km for 
the given QKD scheme
as $\eta=10^{-\frac{\alpha{}l}{10}}$, where $\alpha$ represents 
the loss coefficient of the optical fiber measured in dB/km. The total dB loss of the 
channel is given by $\alpha{}l$.

\subsection{Illustration of the upper bounds}

As discussed in Sec.~\ref{sec_b}, the reduced density matrix of Alice, that we shall denote as 
$\rho_{A_1}^n$, is fixed and cannot be modified by Eve. 
This state has the form 
$\rho_{A_1}^n=\text{Tr}_B(\ket{\varphi_n}_{A_1B}\bra{\varphi_n})=\sum_{k,k'=0}^3 \sqrt{q_kq_{k'}} \bra{n_{k'}}n_k\rangle \ket{k}_{A_1}\bra{k'}$, 
where $\ket{\varphi_n}_{A_1B}$ is given by Eq.~(\ref{eq_nsq}). In the standard BB84 protocol the probabilities 
$q_k$ satisfy $q_k=1/4$. We obtain,
therefore, that $\rho_{A_1}^n$ can be expressed as
\begin{equation}\label{reduced} 
\rho_{A_1}^n = \frac{1}{4} 
\left( \begin{array}{cccc} 
1 & 0 & 2^{-n/2} & 2^{-n/2} \\ 
0 & 1 & 2^{-n/2} & (-1)^n2^{-n/2} \\ 
2^{-n/2} & 2^{-n/2} & 1 & 0 \\
2^{-n/2} & (-1)^n2^{-n/2} & 0 & 1  
\end{array} \right).
\end{equation}
To include this information in the measurement process, we consider that Alice and Bob 
have also access to the results of a set of observables 
$\{C_i\otimes\openone_B\}$ that form a tomographic complete set of Alice's Hilbert space 
$\mathcal{H}_{A_1}$. In particular, we use a Hermitian operator basis
$\{C_1, \ldots, C_{16}\}$. These Hermitian operators satisfy
$\text{Tr}(C_i)=4\delta_{i1}$ and have a Hilbert-Schmidt scalar product $\text{Tr}(C_iC_j)=4\delta_{ij}$.
The probabilities $\text{Tr}(C_i\otimes\openone_B\ \sigma_{A_1B}^{n})=\text{Tr}(C_i\ \rho_{A_1}^n)$, with $\rho_{A_1}^n$ given by Eq.~(\ref{reduced}).

\begin{figure}
\begin{center}
\includegraphics[angle=0,scale=.34]{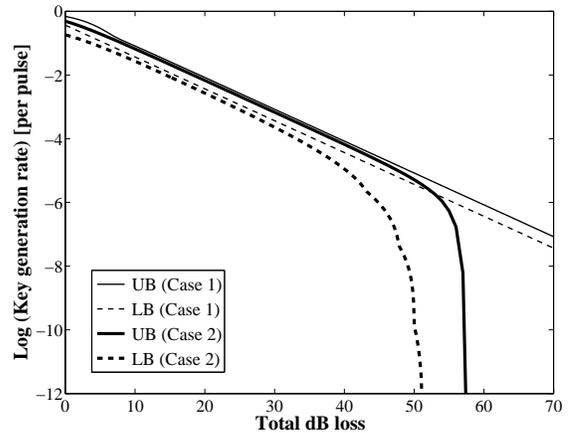}
\end{center}
\caption{Upper bounds on the two-way secret key rate $K$ given by Eq.~(\ref{final_two}) in 
logarithmic scale in comparison to known lower bounds for the same scenario 
given in Ref.~\cite{two_way2}. The figure includes two cases.
(1) No errors in the quantum channel, {\it i.e.}, $Y_0=0$ and $e_{det}=0$. In this case, the upper bound (UB)
is represented by a thin solid line, while the lower bound (LB) is represented by a thin dashed line. (2) 
$Y_0=1.7\times{}10^{-6}$ and $e_{det}=0.033$, 
which correspond to the GYS experiment reported in Ref.~\cite{gys}. In this case, the upper bound (UB)
is represented by a thick solid line, while the lower bound (LB) after 3 B steps is 
represented by a  thick dashed line. 
We assume asymmetric basis choice to suppress the sifting effect \cite{abc}. \label{fig_two_way}}
\end{figure}

\begin{figure}
\begin{center}
\includegraphics[angle=0,scale=.34]{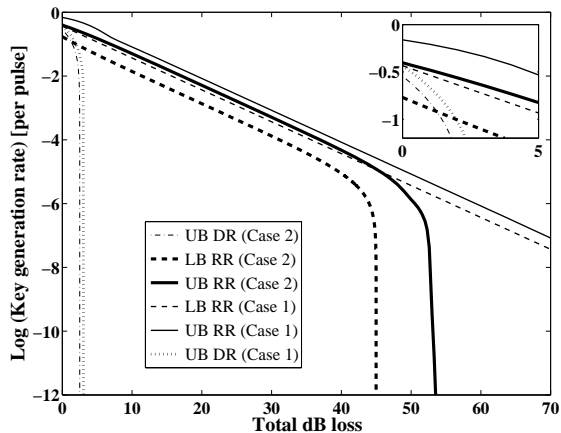}
\end{center}
\caption{Upper bounds on the one-way secret key rate $K_{\rightarrow}$ given by Eq.~(\ref{final_one}) in 
logarithmic scale in comparison to known lower bounds for the same scenario 
given in Refs.~\cite{pracsecproofs1,decoy2}. The figure includes two cases.
(1) No errors in the quantum channel, {\it i.e.}, $Y_0=0$ and $e_{det}=0$. In this case, the upper bound (UB) RR
is represented by a thin
solid line, while the lower bound (LB) is represented by a thin dashed line. (2) 
$Y_0=1.7\times{}10^{-6}$ and $e_{det}=0.033$, 
which correspond to the GYS experiment reported in Ref.~\cite{gys}. In this case, the  
upper bound (UB) RR is represented by a thick solid line, while the lower bound (LB) is represented by a 
thick dashed line. 
The two lines on the left hand side of the graphic represent 
upper bounds for the case of DR (case (1) short dashed line, case (2) dash-dotted line).
The inset figure shows an enlarged view of the upper bounds 
for a total dB loss ranging from 0 to 5 dB.
We assume asymmetric basis choice to suppress the sifting effect \cite{abc}.
\label{fig_one_way}}
\end{figure}

The resulting upper bounds on the two-way and one-way secret key rate  
are illustrated, respectively, in Fig.~\ref{fig_two_way} and Fig.~\ref{fig_one_way}. 
They
state that no secret key can be distilled from the correlations established by the legitimate users 
above the curves, {\it i.e.}, the secret key rate in that region is zero.
These figures include as well {\it lower} bounds for the secret key rate obtained in
Refs.~\cite{pracsecproofs1,decoy2,two_way2}. 
Note, however, the security proofs included in Refs.~\cite{pracsecproofs1,decoy2} 
implicitly assume that Alice and Bob can make
public announcements using two-way communication, and only the error correction and privacy amplification steps 
of the protocol are 
assumed to be realized by means of one-way communication.
We consider the uncalibrated device scenario and
we study two different situations
in each case: (1) no errors in the quantum channel, {\it i.e.}, $Y_0=0$, $e_{det}=0$, 
and (2) $Y_0=1.7\times{}10^{-6}$ and $e_{det}=0.033$. This last scenario corresponds 
to the experimental parameters reported by Gobby-Yuan-Shields (GYS) in  
Ref.~\cite{gys}. Figure~\ref{fig_two_way} and Fig.~\ref{fig_one_way}
do not include the sifting factor of $1/2$ for the BB84 
protocol, since this effect can be avoided by an asymmetric basis choice for 
Alice and Bob \cite{abc}.   
Moreover, we consider that in the asymptotic limit of a large number of transmitted signals most 
of them represent signal states of mean photon number $\mu_0$. That is, the proportion of decoy 
states used to test the behaviour of the quantum channel within the total number of signals sent 
by Alice is neglected. This means that $p_0$ in Eq.~(\ref{rn}) satisfies 
$p_0\approx{}1$ and
\begin{equation}
r_n=\frac{e^{-\mu_0}\mu_0^n}{n!}.
\end{equation}

\subsection{Discussion}

In the case of no errors in the 
quantum channel (Case (1) above) the lower bounds for two-way and one-way QKD derived in 
Refs.~\cite{pracsecproofs1,decoy2,two_way2} coincide.
Furthermore, for low values of the total dB loss, the upper bounds
shown in the figures present a small bump which is specially visible in this last case. 
The origin of this bump is the potential contribution of the 
multi-photon pulses to the key rate.

Let us now consider the cutoff points for 
decoy state QKD in the case 
of errors in the quantum channel (Case (2) above). These are the values of the total dB loss for which the secret key rate drops 
down to zero in Fig.~\ref{fig_two_way} and Fig.~\ref{fig_one_way}. We find that they are given, respectively, by:
$\approx{}51.1$ dB (lower bound two-way after 3 B steps), 
$\approx{}57.4$ dB (upper bound two-way),
$\approx{}44.9$ dB (lower bound one-way), and
$\approx{}53.5$ dB (upper bound one-way with RR). These quantities can be related 
with the following transmission distances:  $179.2$ km, $209.2$ km, $149.6$ km and $190.6$ km.
Here we have used $\alpha=0.21$ dB/km and the efficiency of Bob's detectors is $4.5\%$ \cite{gys}.
It is interesting to compare the two-way cutoff point of $209.2$ km with a similar distance 
upper bound of $208$ km provided in Ref.~\cite{two_way2} for the same values of the 
experimental parameters. Note, however, that the upper bound derived in Ref.~\cite{two_way2} relies on the assumption
that a secure key can only be extracted from single photon states. That is, it 
implicitly assumes the standard BB84 protocol. If this 
assumption is removed and one also includes in the analysis the 
potential contribution of the multi-photon signals to the key rate (due, for instance, to the SARG04 protocol \cite{sarg}),
then the cutoff point provided in Ref.~\cite{two_way2}
transforms from $208$ km 
to $222$ km, which is above 
the $209.2$ km presented here.

Figure~\ref{fig_one_way} shows a significant difference between 
the behaviour of
the upper bounds for one-way classical post-processing with RR 
and DR. Most importantly, the upper bounds on 
$K_{\rightarrow}$ for the case of DR can be below the lower bounds 
on the secret key rate 
derived in Refs.~\cite{pracsecproofs1,decoy2}.
Note, however, that the scenario considered here is  slightly 
different from the one assumed
in the security proofs of Refs.~\cite{pracsecproofs1,decoy2}. 
In particular, the analysis contained in 
Sec.~\ref{sec_one_way} for the case of DR
does not allow {\it any} communication from Bob to Alice
once the conditional probabilities $p_{kj}^n$ are determined.
This means, for instance, that Bob cannot even declare in which particular 
events his detection apparatus produced a 
``click". However, as mentioned previously,
Refs.~\cite{pracsecproofs1,decoy2} 
implicitly assume that only 
the error correction and privacy amplification steps 
of the protocol are performed with one-way communication.
If the analysis performed in Sec.~\ref{sec_one_way} is modified such that
Bob 
is now allowed to inform Alice
which signal states he actually detected, 
then it turns out that the resulting 
upper bounds 
in this modified scenario coincide with 
those derived for the case of RR. To include this initial  
communication step from Bob to Alice 
in the analysis, one can 
use the following procedure.
Let the projector $\Pi_{A_1B}$ be defined as
\begin{equation}
\Pi_{A_1B}=\openone_{A_1}\otimes(\openone_B-\ket{vac}_B\bra{vac}). 
\end{equation}
Then, 
one can add to Eq.~(\ref{primalSDP_one_way}) one
extra constraint 
\begin{equation}\label{hoy_playa}
\sigma_{A_1B}^{n\ post}({\bf x})=\frac{\Pi_{A_1B}\sigma_{A_1B}^n({\bf x})\Pi_{A_1B}}{Y_n},
\end{equation} 
and substitute the condition $\sigma_{A_1B}^n({\bf x})-\sigma_{ext}^n({\bf x})\geq{}0$ by
\begin{equation}\label{hoy_playa2} 
\sigma_{A_1B}^{n\ post}({\bf x})-\sigma_{ext}^n({\bf x})\geq{}0.
\end{equation}
Equation~(\ref{hoy_playa}) refers to 
the normalized state that is postselected by Alice and Bob once Bob declares 
which signals he detected. Equation~(\ref{hoy_playa2}) indicates that 
the BEA has to be applied to this postselected state.
Finally, each term in the summation 
given by Eq.~(\ref{final_one}) has to be multiplied by the yield $Y_n$, {\it i.e.}, the probability that 
Bob obtains a ``click" conditioned on the fact that Alice sent an $n$-photon state. 

Our numerical results 
indicate that the 
upper bounds given by Eq.~(\ref{final_two}) and Eq.~(\ref{final_one}) 
are close to the known lower bounds available in the 
scientific literature for the same scenarios. However, one might expect that these upper bounds 
can be further tightened 
in different ways. For instance, by substituting in Eq.~(\ref{final_two}) and Eq.~(\ref{final_one}) 
the Shannon mutual information with any other tighter 
upper bound on the secret key rate that can be extracted
from a classical tripartite probability distribution measured on a purification of the 
state $\rho_{ent}^{n}$ (in the case of two-way 
QKD) or of the state $\rho_{ne}^{n}$ (one-way QKD). Moreover, as they are, 
Eq.~(\ref{final_two}) and Eq.~(\ref{final_one})
implicitly assume that the legitimate users 
know precisely the number 
of photons contained in each signal emitted. 
However, in decoy state QKD Alice and Bob have only access to the 
conditional joint probability distribution describing their outcomes given that Alice emitted 
an $n$-photon state, but they do not have single shot photon number resolution of each 
signal state sent. 

As a side remark, we would like to emphasize that to calculate the upper bounds given by
Eq.~(\ref{final_two}) and Eq.~(\ref{final_one}) it is typically sufficient to consider only a finite 
number of terms in the summations. This result arises from 
the limit imposed by the unambiguous state discrimination (USD)
attack \cite{usda}. This attack does not introduce any errors in Alice's and Bob's signal states. 
Moreover, it corresponds to an entanglement-breaking channel 
\cite{ebc} and, therefore, it cannot lead to a secure key both for 
the case of two-way and one-way QKD
\cite{curty04a,tobione}. The maximum probability of unambiguously discriminating
an $n$-photon state sent by Alice is given by \cite{usda}
\begin{equation} 
P_{D}^n = \left\{ \begin{array}{ll} 
0 & \textrm{$n\leq{}2$}\\ 
1-2^{1-n/2} & \textrm{$n$ even}\\ 
1-2^{(1-n)/2} & \textrm{$n$ odd}. 
\end{array} \right. 
\end{equation}
For typical observations this quantity can be related 
with a transmission efficiency $\eta_n$ of the quantum channel, {\it i.e.}, an $\eta_n$ 
that provides an expected click rate at Bob's side equal to $P_D^n$. This last condition can be written 
as 
\begin{equation}
\eta_n=1-(1-P_D^n)^{1/n}.
\end{equation}
Whenever the overall transmission probability of each photon satisfies $\eta\leq{}\eta_n$, then any pulse 
containing $n$ or more photons is insecure against the USD attack. After a short calculation, we obtain 
that the total number of $n$-photon signals that need to be considered in the summations of Eq.~(\ref{final_two}) and 
Eq.~(\ref{final_one}) can be upper bounded as 
\begin{equation} 
n \leq \left\{ \begin{array}{ll} 
\big\lfloor\frac{1}{\log_2[\sqrt{2}(1-\eta)]}\big\rfloor & \textrm{$n$ even}\\ 
\big\lfloor\frac{1}{2\log_2[\sqrt{2}(1-\eta)]}\big\rfloor & \textrm{$n$ odd}. 
\end{array} \right. 
\end{equation}

\section{Conclusion}\label{CONC}

In this paper we have derived
upper bounds on the secret key rate and distance that 
can be covered by two-way and one-way 
decoy state quantum key distribution (QKD). Our analysis considers the uncalibrated 
device scenario and we have assumed the typical initial post-processing step where double click events 
are randomly assigned to single click events.
We have used two preconditions for secure two-way and one-way QKD. 
In particular, the legitimate users need to prove that there exists no separable state (in the case of two-way 
QKD), or that there exists no quantum state having a symmetric extension (one-way QKD), 
that is compatible with the available measurements results. Both criteria have been previously employed 
in the scientific literature to 
evaluate single-photon implementations of QKD. Here we have applied them 
to investigate a realistic source of weak coherent pulses,
and
we have shown that they can be formulated as a convex optimization 
problem known as a semidefinite program (SDP).  
Such instances of convex optimization problems can 
be solved efficiently, for example by means of the interior-point methods.

As a result, we have obtained fundamental limitations on the performance of 
decoy state QKD when this initial post-processing of the double clicks is performed.
These upper bounds cannot be overcome by any classical 
communication technique
(including, for example, SARG04 protocol, adding noise protocols, degenerate codes and two-way classical post-processing protocols)
that the legitimate users may employ to process their correlated data afterwards.
Moreover, our
results seem to be already close to well known lower bounds for the same scenarios,
thus showing that there are clear limits to the further improvement of classical post-processing techniques
in decoy state QKD.

The analysis 
presented in this paper could as well be straightforwardly adapted to evaluate
other implementations of the BB84 protocol with practical signals like, for example, 
those experimental demonstrations based on  WCP without 
decoy states or on entangled signals coming from a parametric down conversion source.

\section{ACKNOWLEDGEMENTS}

M.C. especially thanks H.-K. Lo and 
N. L\"utkenhaus 
for hospitality and support during his stay at the University of Toronto and at the 
Institute for Quantum Computing (University of Waterloo) 
where this manuscript was finished. This work was supported by the European Projects 
SECOQC and QAP, NSERC, Quantum Works, CSEC, 
CFI, CIPI, CIFAR, the CRC program, MITACS, OIT, OCE, 
by Xunta de Galicia (Spain, Grant No. INCITE08PXIB322257PR), and by University 
of Vigo (Program ``Axudas \'a mobilidade dos investigadores").

\bibliographystyle{apsrev}

\begin{thebibliography}{99}
\bibitem{nwnt} Here we define quantum correlations as those correlations that cannot be explained by means of an 
intercept-resend attack \cite{curty04a,curty04b}.
\bibitem{initial_2w}
Before restricting Alice and Bob to only communicate one-way, one typically allows them to realize an initial 
two-way communication step 
to estimate the joint probability distribution describing their measurements outcomes.  This is the approach that we 
follow in this paper. In decoy state QKD this is also important in order to distinguish between signal and decoy pulses and to 
estimate the conditional probabilities $p_{kj}^n$.
\bibitem{Note_sym} From now on, the term extension will always stand for a symmetric extension to 
two copies of system $A$ or $B$. We will not make any further distinction between 
the different types of extension and we simply call the state extendible. The extension to two copies 
of system $B$ corresponds to DR, and extensions to two copies of system $A$ 
corresponds to RR.
\end{thebibliography}

\end{document}